\documentclass[twocolumn,showpacs,prb,superscriptaddress,floatfix]{revtex4}

\usepackage{color} 
\usepackage{tabularx} 
\usepackage{epsfig}
\usepackage{amsmath} 
\usepackage{amssymb} 
\usepackage{graphicx}
\usepackage{wasysym}
\usepackage{oldgerm}
\usepackage{psfrag}

\newcommand{\chisg}{\chi_{_{\rm SG}}}

\begin{document}

\title{Spin glasses in a field: Three and four dimensions as seen from
one space dimension}

\author{Derek Larson}
\affiliation{Department of Physics, National Taiwan University, No.~1,
Sec.~4, Roosevelt Rd., Taipei 106, Taiwan}

\author{Helmut G.~Katzgraber}
\affiliation{Department of Physics and Astronomy, Texas A\&M University,
College Station, Texas 77843-4242, USA}
\affiliation{Theoretische Physik, ETH Zurich, CH-8093 Zurich,
Switzerland}

\author{M.~A.~Moore}
\affiliation{School of Physics and Astronomy, University of Manchester,
Manchester M13 9PL, United Kingdom}

\author{A.~P.~Young}
\affiliation{Department of Physics, University of California, Santa Cruz,
California 95064, USA}

\date{\today}

\begin{abstract}

We study the existence of a line of transitions of an Ising spin glass
in a magnetic field---known as the de Almeida-Thouless line---using
one-dimensional power-law diluted Ising spin-glass models. We choose the
power-law exponent to have values that approximately correspond to
three- and four-dimensional nearest-neighbor systems and perform a
detailed finite-size scaling analysis of the data for large linear
system sizes, using both a new approach proposed recently
[Phys.~Rev.~Lett.~{\bf 103}, 267201 (2009)], as well as traditional
approaches. Our results for the model corresponding to a
three-dimensional system are consistent with there being no de
Almeida-Thouless line, although the new finite-size scaling approach
does not rule one out. For the model corresponding to four space
dimensions, the new and traditional finite-size scaling analyses give
conflicting results, indicating the need for a better understanding of
finite-size scaling of spin glasses in a magnetic field.

\end{abstract}

\maketitle

\section{Introduction}
\label{sec:intro}

One of the most striking predictions of the mean-field theory of Ising
spin glasses, taken to be the exact solution\cite{parisi:80} of the
infinite-range Sherrington-Kirkpatrick (SK)\cite{sherrington:75} model,
is the existence of a line of transitions, known as the de
Almeida-Thouless\cite{almeida:78} (AT) line in the presence of a
magnetic field. This line of transitions separates a high-temperature
region where the description of the model is quite simple, just
involving a single order parameter, from a low-temperature region where
there is ``replica symmetry breaking'' (RSB) in which the system has an
infinite number of order parameters characterized by a
function.\cite{parisi:80}

The question of whether RSB applies to realistic short-range spin
glasses remains controversial. According to the RSB picture, real spin
glasses behave rather similarly to the SK model and so have an AT line.
However, according to the phenomenological ``droplet''
picture\cite{fisher:87,fisher:88,bray:86,mcmillan:84a} there is no AT
line and the zero-field transition is rounded out by any nonzero
magnetic field in finite-dimensional short-range systems.

It is convenient for simulations that a static property, the spin-glass
susceptibility $\chisg$, diverges at the transition. This quantity is
the inverse of the eigenvalue of the stability matrix\cite{almeida:78}
found by de Almeida and Thouless and is given by the zero-wave-vector
$k=0$ limit of $\chisg(k)$ defined in Eq.~\eqref{eq:chisg} below.
Because $\chisg$ can be computed in simulations
directly,\cite{experiment} one might imagine that it would be
straightforward to decide if the AT line occurs in, say, a
three-dimensional spin glass. However, there is still no consensus on
this issue because there seem to be quite large corrections to
finite-size scaling (FSS). The purpose of this paper is to investigate
different methods that have been proposed to perform FSS to see if there
is an AT line in three and in four space dimensions.

In fact, rather than to study short-range
models,\cite{ciria:93b,migliorini:98,marinari:98d,houdayer:99,krzakala:01,young:04,takayama:04,joerg:08a}
we find it convenient to study a one-dimensional model with long-range
interactions which is taken as a \textit{proxy} for a short-range
model.\cite{katzgraber:03,katzgraber:09b,larson:10,banos:12b,leuzzi:08,leuzzi:09,leuzzi:11}
The interactions of the long-range model fall off with a power $\sigma$
of the distance and varying the power effectively corresponds
to varying the space dimension of the corresponding short-range
model. Here we study long-range models which are proxies for
short-range models in three and four space dimensions.

The outline of this paper is as follows. Section \ref{sec:model}
describes the model, the quantities we calculate and details of the
numerical simulations. Section \ref{sec:results} describes the results,
and Sec.~\ref{sec:conclusions} gives our conclusions.

\section{Model, Observables \& Numerical Details}
\label{sec:model}

\subsection{Model}

We study a variation of the model introduced in
Ref.~\onlinecite{leuzzi:08}, which is given by the Hamiltonian
\begin{equation}
{\mathcal H} = 
-\sum_{i,j} \varepsilon_{ij} J_{ij} S_i S_j - \sum_i h_i S_i\; .
\label{eq:hamiltonian}
\end{equation}
In Eq.~\eqref{eq:hamiltonian}, $S_i = \pm 1$ are Ising spins placed on a
ring of length $L$ to enforce periodic boundary conditions in a natural
way.\cite{katzgraber:03} The interactions $J_{ij}$ are chosen from a
Gaussian distribution with zero mean and standard deviation unity. The
dilution matrix $\varepsilon_{ij}$ takes values $0$ and $1$, and has the
probability $p_{ij} \sim r_{ij}^{-2\sigma}$ of taking the value $1$,
where $r_{ij} = (L/\pi)\sin(\pi|i - j|/L)$ is the geometric distance
between the spins. To prevent the probability of placing a bond between
two spins being larger than $1$, a short-distance cutoff is applied and,
thus, we take
\begin{equation}
p_{ij} = 1 - \exp(-{\mathcal A}/r_{ij}^{2\sigma}) \, .
\label{eq:prob}
\end{equation}
The constant ${\mathcal A}$ is determined by the requirement that the
\textit{mean} coordination number, $z_\text{av}$, takes a specified
value
\begin{equation}
z_\text{av} = \sum_{i = 1}^{L - 1} p_{iL} ,
\end{equation}
and we set $z_\text{av}=6$. The values of ${\mathcal A}$ are given in
Table \ref{tab:simparams}. The site-dependent random fields $h_i$ are
chosen from a Gaussian distribution with zero mean and standard
deviation $H$.

By tuning the exponent $\sigma$ in Eq.~\eqref{eq:prob} one can change
the universality class of the model in Eq.~\eqref{eq:hamiltonian} from
the infinite-range to the short-range universality case. For $0 <
\sigma \le 1/2$ the model is in the infinite-range universality
class\cite{mori:11,wittmann:12} and, in particular, $\sigma = 0$
corresponds to the Viana-Bray model.\cite{viana:85} For $1/2 < \sigma
\le 2/3$ the model describes a mean-field, long-range spin
glass,\cite{katzgraber:03} corresponding to a short-range model with a
space dimension above the upper critical dimension, i.e.,~$d \ge d_{\rm
u} = 6$.\cite{harris:76,katzgraber:08} For $2/3 < \sigma \le 1$ the
model has non-mean-field critical behavior with a finite transition
temperature $T_c$, while for $\sigma \ge 1$, the transition temperature
is zero.\cite{katzgraber:03} The value of $\sigma$ in the long-range
one-dimensional model corresponds roughly to an effective space
dimension $d$ in a short-range model via the
relation\cite{katzgraber:09b,larson:10,banos:12b}
\begin{equation}
d = \frac{2 - \eta(d)}{2 \sigma - 1} \, ,
\label{eq:d_sigma}
\end{equation}
where $\eta(d)$ is the critical exponent $\eta$ for the short-range
model, which is zero in the mean-field regime. Here we are interested in
values of $\sigma$ that correspond to three- (3D) and four-dimensional
(4D) short-range systems. Because $\eta(d = 3) =
-0.384(9)$,\cite{hasenbusch:08} three dimensions corresponds to
$\sigma_{d = 3} \simeq 0.896$, and since $\eta(d = 4) =
-0.275(25)$,\cite{joerg:08d} four space dimensions corresponds to
$\sigma_{d = 4} \simeq 0.784$.

\subsection{Observables}

To determine whether a spin-glass state exists in a magnetic field, we
study the wave-vector-dependent spin-glass susceptibility defined by
\begin{equation} 
\chisg(k) = \frac{1}{L} \sum_{i, j} 
\left[\Big( 
\langle S_i S_j\rangle_T - \langle S_i \rangle_T \langle S_j\rangle_T 
\Big)^2 \right]_{\rm av}\!\!\!\!\! e^{ik\, (i-j)} , 
\label{eq:chisg} 
\end{equation} 
where $\langle \cdots \rangle_T$ denotes a thermal average and
$[\cdots]_{\rm av}$ an average over the disorder. Each thermal average
is obtained from a separate spin replica, i.e., we simulate four copies
with the same disorder but different Markov chains at each temperature.
As discussed in Sec.~\ref{sec:intro}, $\chisg(k)$ is an appropriate
quantity to study because $\chisg \equiv \chisg(k=0)$ diverges on the AT
line.

\begin{figure*}[!tbh]
\center

\includegraphics[width=0.45\textwidth]{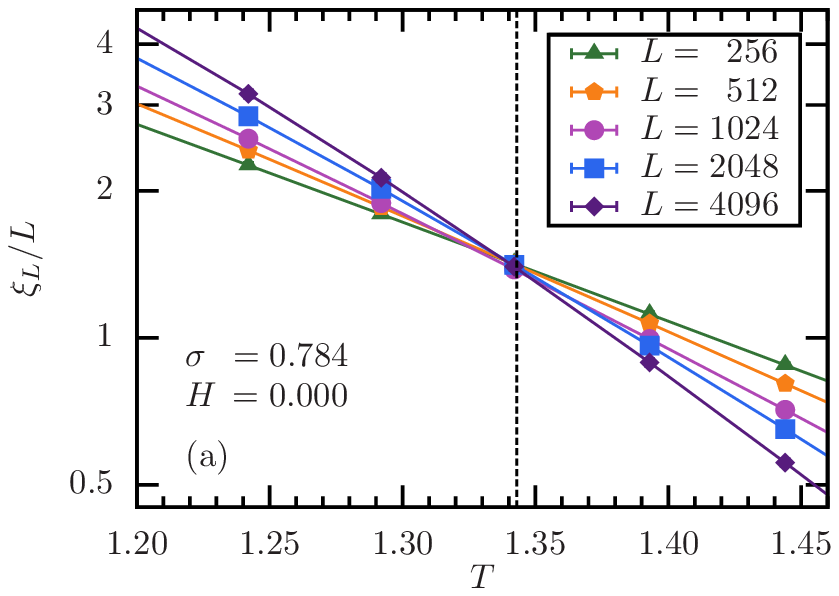}
\hspace*{0.5cm}
\includegraphics[width=0.45\textwidth]{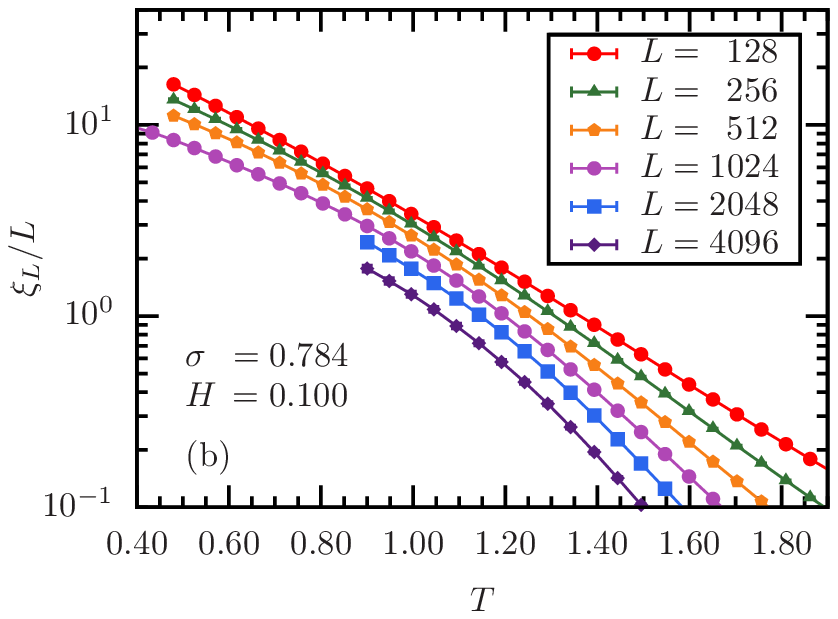}

\vspace*{0.1cm}

\includegraphics[width=0.45\textwidth]{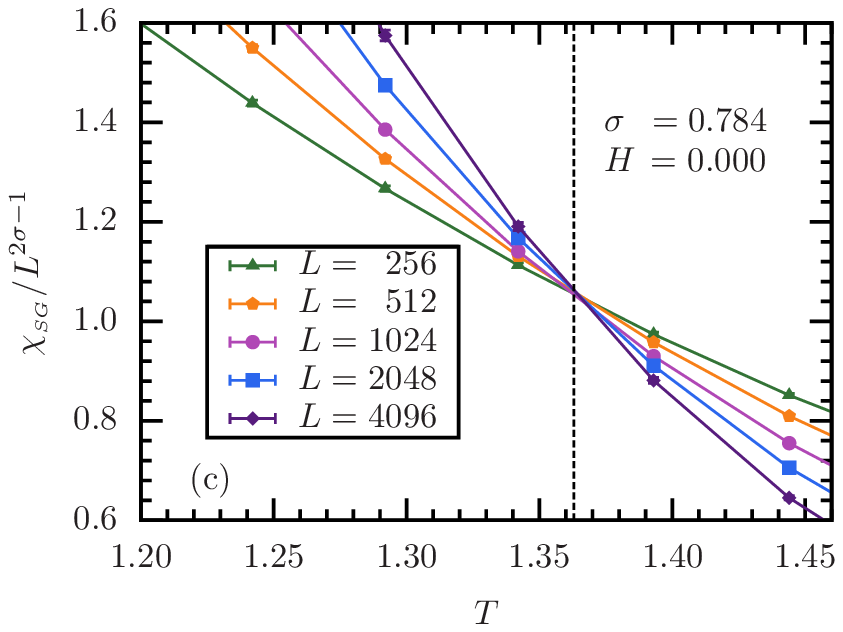}
\hspace*{0.5cm}
\includegraphics[width=0.45\textwidth]{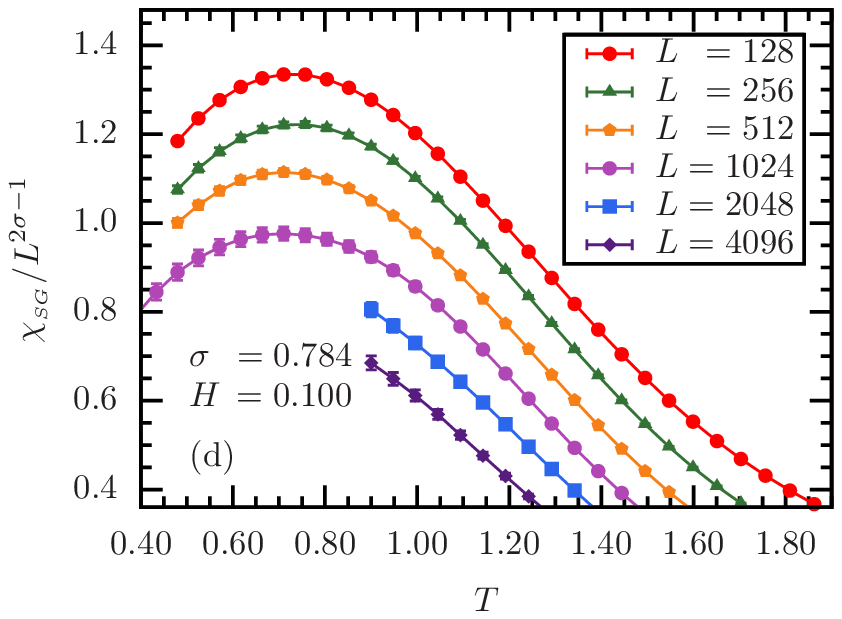}

\vspace*{-0.3cm}
\caption{(Color online)
Data used in a standard FSS for $\sigma = 0.784$, which is a proxy
for four dimensions. The left column is for $H = 0$ and the right column
for $H = 0.1$. The top row shows the finite-size correlation length
divided by $L$ as a function of temperature $T$ for different system
sizes. The bottom row shows results for the scale-invariant spin-glass
susceptibility, $\chisg/L^{2\sigma-1}$. The intersections seen in the
left-hand column indicate a transition in zero field. The intersection
temperatures (shown by the dashed vertical line) are $T_c \simeq 1.34$
from $\xi_L/L$ data in (a), and $T_c \simeq 1.36$ from $\chi_{\rm
SG}/L^{2\sigma - 1}$ data in (c). The small difference is likely
due to sensitive corrections to scaling in the susceptibility (see
Ref.~\onlinecite{katzgraber:06} for details). By contrast, the lack
of intersections in the right-hand column indicates no transition in
a magnetic field.
}
\label{fig:0.784-standard}
\end{figure*}

It is also convenient to extract from the spin-glass susceptibility a
correlation length, which is usually defined by
\cite{palassini:99b,ballesteros:00,young:04,katzgraber:09b}
\begin{equation}
\xi_L = \frac{1}{2 \sin (k_{1}/2)}
\left[
\frac{\chisg(0)}{\chisg(k_{1})} - 1
\right]^{1/(2\sigma-1)} ,
\label{eq:xiL}
\end{equation}
where $k_{1} = 2 \pi / L$ is the smallest nonzero wave vector. Because
we work in the non-mean-field regime, standard finite-size scaling (FSS)
applies, i.e., 
\begin{equation}
{\xi_L}/{L} = {\mathcal X} [ L^{1/\nu} (T - T_c) ] \, .
\label{eq:xiscale}
\end{equation}
The importance of $\xi_L/L$ is that it is dimensionless.  As such,
data for ${\xi_L}/{L}$ for different system sizes $L$ cross at the
transition temperature $T=T_c$ if corrections to FSS are unimportant
[see, for example, Fig.~\ref{fig:0.784-standard}(a)]. This is a
particularly convenient way to locate $T_c$.

For the one-dimensional model the critical exponent $\eta_{\rm LR}$
satisfies the \textit{exact} relation $2 - \eta_{\rm LR} = 2 \sigma -
1$, so it is also convenient to study the finite-size scaling of a
second scale-invariant quantity, namely
\begin{equation}
\chisg / L^{2-\eta_{\rm LR}} = 
\chisg/L^{2\sigma-1} =
{\mathcal C}[L^{1/\nu} (T - T_c)] \, .
\label{eq:chiscale}
\end{equation}
Again, if FSS applies without corrections, data for
$\chisg/L^{2\sigma-1}$ for different system sizes cross at $T_c$.

\begin{table}
\caption{
Parameters of the simulations for different field strengths $H$
and exponents $\sigma$. $N_{\rm sa}$ is the number of samples, $N_{\rm
sw}$ is the total number of Monte Carlo sweeps, $T_{\rm min}$ is the
lowest temperature simulated, and $N_T$ is the number of temperatures
used in the parallel tempering method for each system size $L$. The
last column shows the parameter ${\mathcal A}$ in Eq.~\eqref{eq:prob} for 
$z_{\rm av} = 6$ neighbors.
\label{tab:simparams}}
{\footnotesize
\begin{tabular*}{\columnwidth}{@{\extracolsep{\fill}} c r r r r r r r}
\hline
\hline
$\sigma$ & $H$ & $L$ & $N_{\rm sa}$ & $N_{\rm sw}$ & $T_{\rm min}$ &
$N_{T}$ & ${\mathcal A}$
\\
\hline

$0.784$ & $0.0$ & $ 256$ & $10179$ & $20$ & $0.617$ & $43$ & $1.86026$ \\
$0.784$ & $0.0$ & $ 512$ & $10068$ & $20$ & $0.617$ & $43$ & $1.82182$ \\
$0.784$ & $0.0$ & $1024$ & $10240$ & $20$ & $0.617$ & $43$ & $1.79706$ \\
$0.784$ & $0.0$ & $2048$ & $ 4810$ & $20$ & $0.710$ & $17$ & $1.78084$ \\
$0.784$ & $0.0$ & $4096$ & $ 4400$ & $20$ & $1.192$ & $17$ & $1.77010$ \\[2mm]

$0.784$ & $0.1$ & $ 128$ & $37484$ & $20$ & $0.480$ & $46$ & $1.92172$ \\
$0.784$ & $0.1$ & $ 192$ & $30150$ & $20$ & $0.480$ & $46$ & $1.88220$ \\
$0.784$ & $0.1$ & $ 256$ & $37494$ & $20$ & $0.480$ & $46$ & $1.86026$ \\
$0.784$ & $0.1$ & $ 384$ & $27970$ & $20$ & $0.480$ & $46$ & $1.83571$ \\
$0.784$ & $0.1$ & $ 512$ & $32856$ & $20$ & $0.480$ & $46$ & $1.82182$ \\
$0.784$ & $0.1$ & $ 768$ & $ 9988$ & $20$ & $0.344$ & $49$ & $1.80607$ \\
$0.784$ & $0.1$ & $1024$ & $ 9995$ & $21$ & $0.344$ & $49$ & $1.79706$ \\
$0.784$ & $0.1$ & $1536$ & $ 7163$ & $20$ & $0.900$ & $28$ & $1.78676$ \\
$0.784$ & $0.1$ & $2048$ & $ 5116$ & $20$ & $0.900$ & $28$ & $1.78084$ \\
$0.784$ & $0.1$ & $3072$ & $ 4306$ & $20$ & $0.900$ & $28$ & $1.77403$ \\
$0.784$ & $0.1$ & $4096$ & $ 6592$ & $20$ & $0.900$ & $28$ & $1.77010$ \\[2mm]

$0.896$ & $0.0$ & $ 128$ & $14800$ & $20$ & $0.617$ & $43$ & $2.78392$ \\
$0.896$ & $0.0$ & $ 192$ & $10770$ & $20$ & $0.617$ & $43$ & $2.76125$ \\
$0.896$ & $0.0$ & $ 256$ & $ 9830$ & $20$ & $0.617$ & $43$ & $2.74955$ \\
$0.896$ & $0.0$ & $ 512$ & $11632$ & $20$ & $0.617$ & $43$ & $2.73087$ \\
$0.896$ & $0.0$ & $1024$ & $11052$ & $20$ & $0.617$ & $43$ & $2.72041$ \\
$0.896$ & $0.0$ & $2048$ & $ 4345$ & $20$ & $0.617$ & $43$ & $2.71446$ \\
$0.896$ & $0.0$ & $4096$ & $ 5230$ & $20$ & $0.710$ & $17$ & $2.71105$ \\[2mm]

$0.896$ & $0.1$ & $ 128$ & $45000$ & $20$ & $0.300$ & $50$ & $2.78392$ \\
$0.896$ & $0.1$ & $ 256$ & $32333$ & $20$ & $0.480$ & $46$ & $2.74955$ \\
$0.896$ & $0.1$ & $ 384$ & $45000$ & $20$ & $0.300$ & $50$ & $2.73732$ \\
$0.896$ & $0.1$ & $ 512$ & $28603$ & $20$ & $0.480$ & $46$ & $2.73087$ \\
$0.896$ & $0.1$ & $ 768$ & $ 8036$ & $21$ & $0.480$ & $46$ & $2.72405$ \\
$0.896$ & $0.1$ & $1024$ & $ 9285$ & $21$ & $0.300$ & $50$ & $2.72041$ \\

\hline
\hline
\end{tabular*}
}
\end{table}

Both the correlation length in Eq.~\eqref{eq:xiscale} and the spin-glass
susceptibility in Eq.~\eqref{eq:chiscale} involve $k=0$ fluctuations.
Recently, Refs.~\onlinecite{leuzzi:09}, \onlinecite{leuzzi:11}, and
\onlinecite{banos:12} have argued that one should avoid data at $k=0$
for spin glasses in the presence of a magnetic field on the grounds that
there are large corrections to FSS. We discuss this in detail below and
for now just present the new proposed quantities to be
measured\cite{leuzzi:09,leuzzi:11,banos:12} that avoid $k=0$
fluctuations. We shall use the term ``modified'' FSS analysis
to denote the use of these quantities. Below we compare the results of
this modified analysis with results obtained from
Eqs.~\eqref{eq:xiscale} and \eqref{eq:chiscale}, which we denote the
``standard'' FSS approach.

At long wavelength one expects
\begin{equation}
\chisg(k)^{-1} = \chisg(0)^{-1} + By + Cy^2
\label{eq:parfit}
\end{equation}
with $y = k^{2\sigma-1}$, which is the generalization of the
Ornstein-Zernicke equation to long-range interactions, so one can
calculate $\chisg(0)^{-1}$ \textit{indirectly} by fitting data for
nonzero $k$ to this form. For $L \to \infty$, this extrapolated
value, $\chisg(k\to 0)^{-1}$, vanishes at and below the transition
temperature.  Interestingly, one finds\cite{leuzzi:09,leuzzi:11}
that the extrapolated value goes through zero even for a finite
system, at a temperature $T^\star(L)$ which tends to $T_c$ for $L \to
\infty$. This means $T_c = \lim_{L \to \infty} T^\star(L)$. We find
quite strong corrections to the asymptotic result (as found also for
the Sherrington-Kirkpatrick model in Ref.~\onlinecite{billoire:03}),
so we include the leading correction to scaling by fitting the data to
\begin{equation}
T^\star(L) = T_c + {A \over L^\lambda},
\label{eq:Tc}
\end{equation}
where $\lambda$ is a correction to scaling exponent and $A$ is the
amplitude of the correction.

\begin{figure*}[!tbh]
\center
\includegraphics[width=0.45\textwidth]{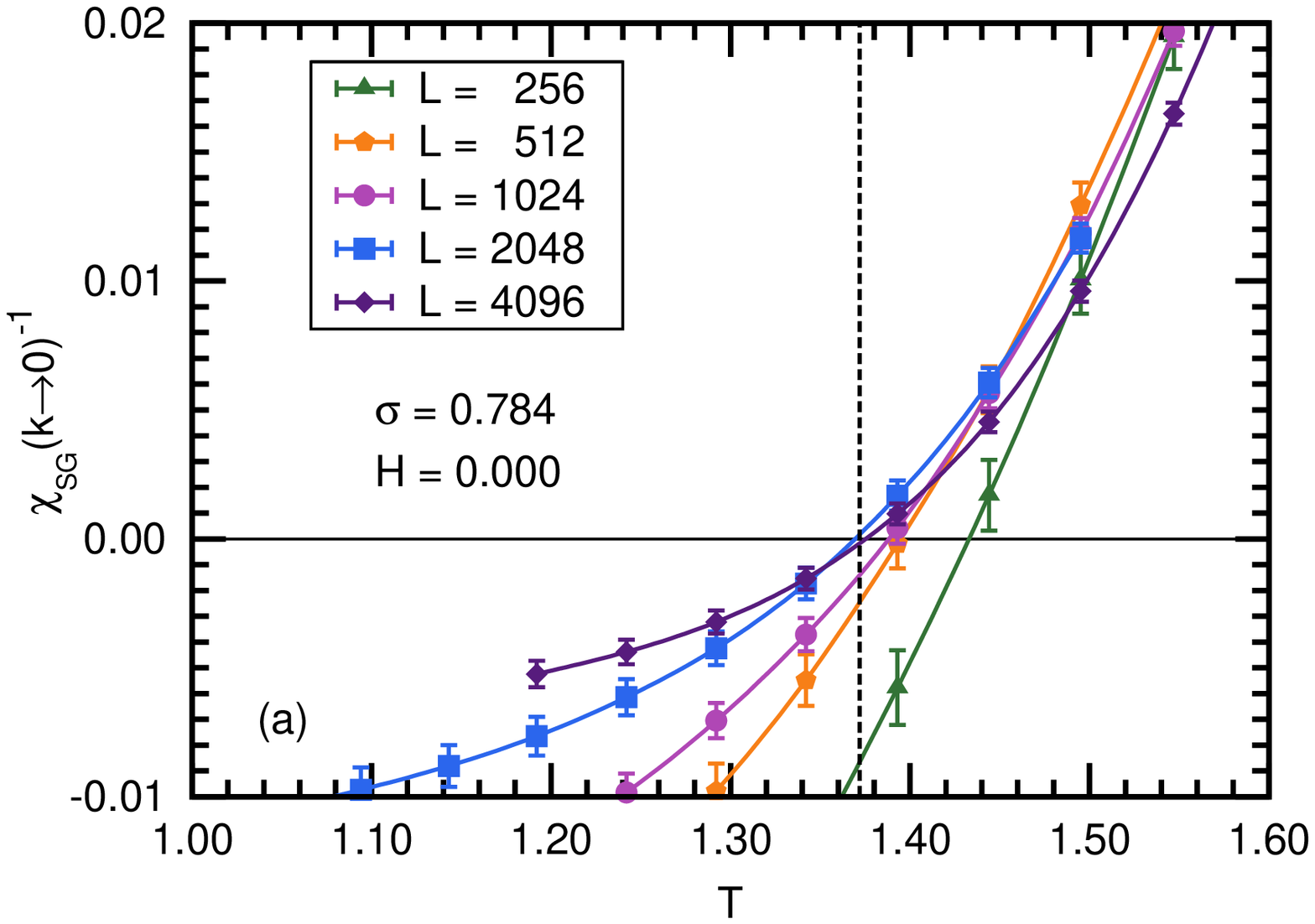}
\hspace*{0.5cm}
\includegraphics[width=0.45\textwidth]{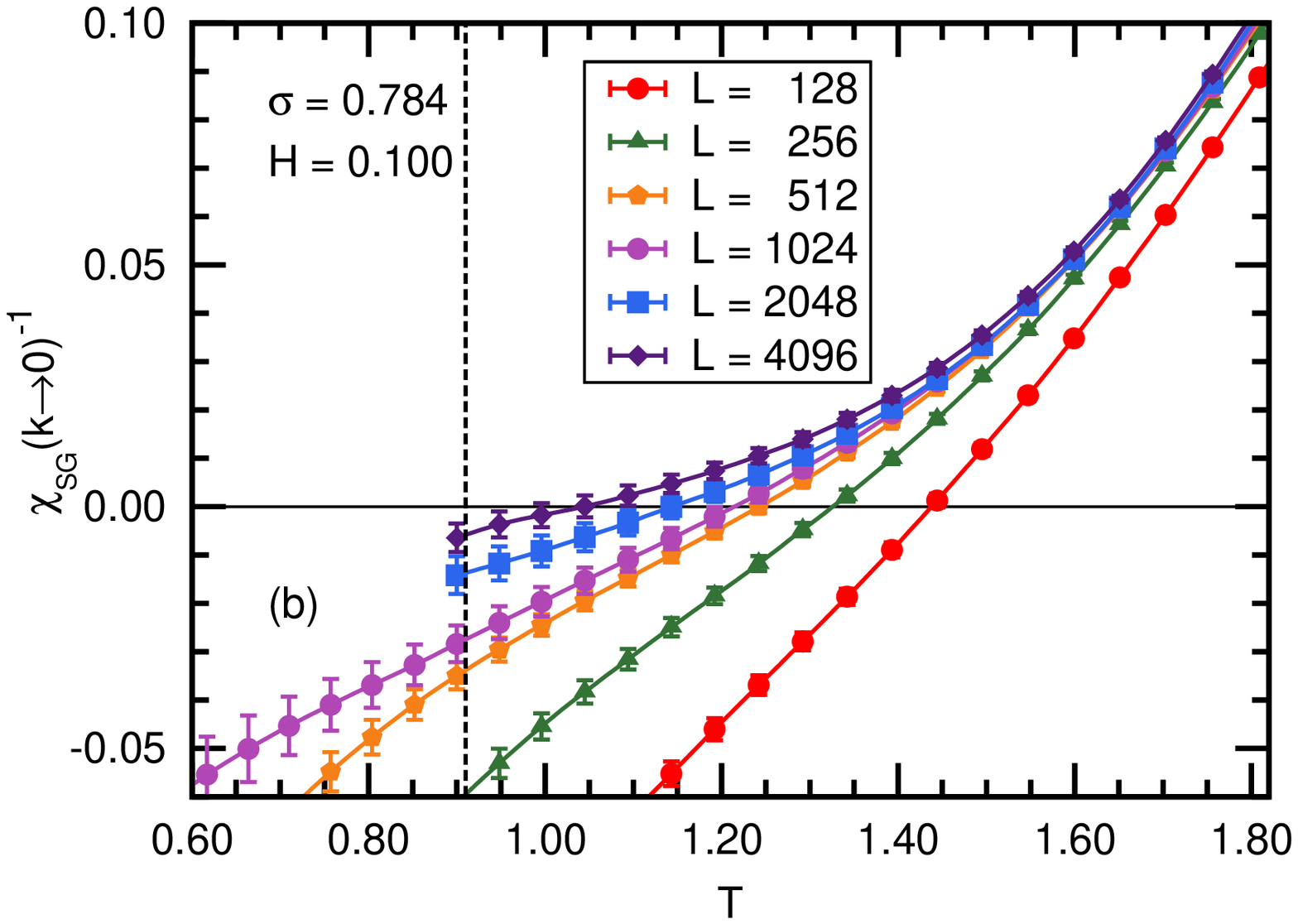}

\vspace*{0.1cm}

\includegraphics[width=0.45\textwidth]{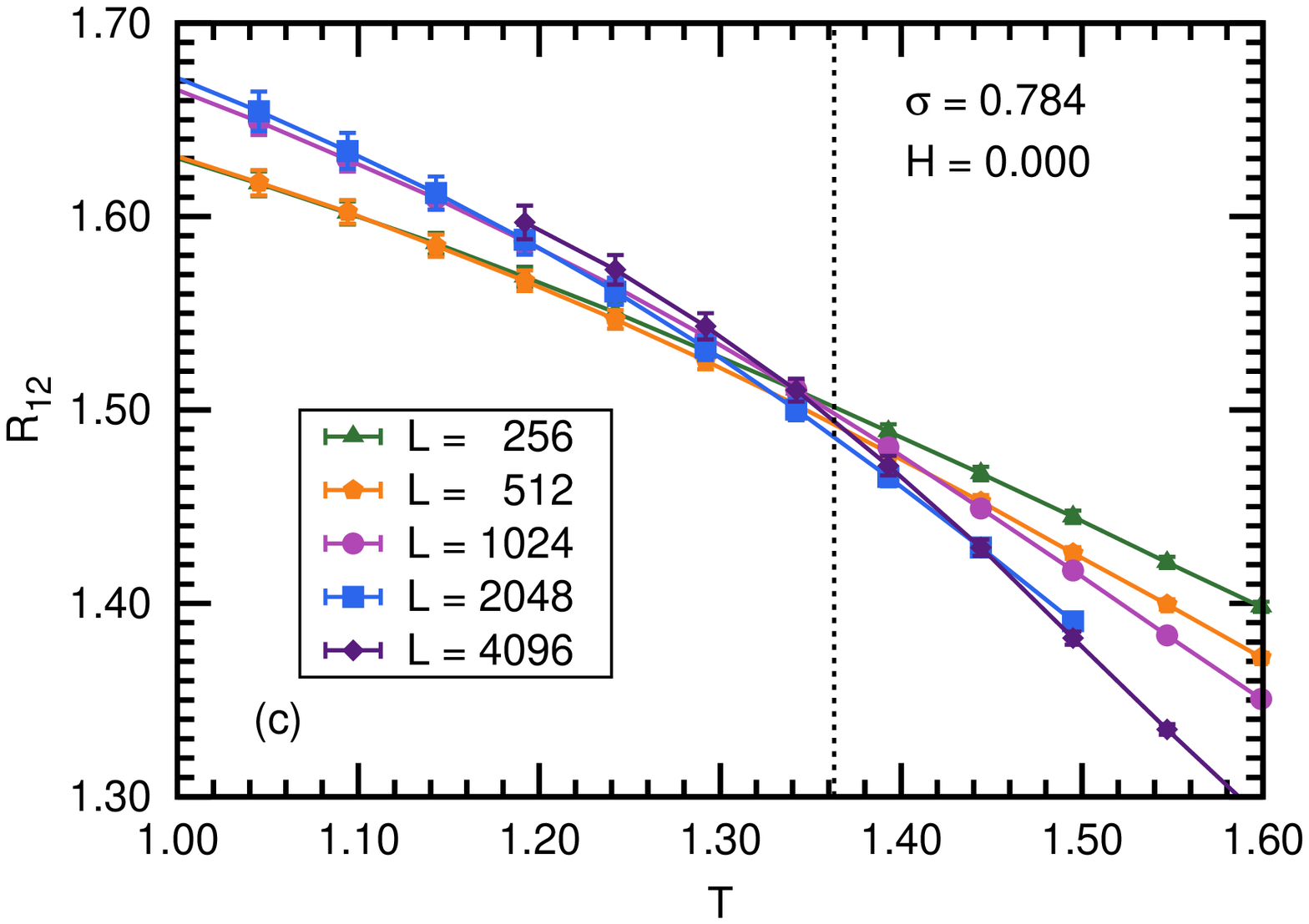}
\hspace*{0.5cm}
\includegraphics[width=0.45\textwidth]{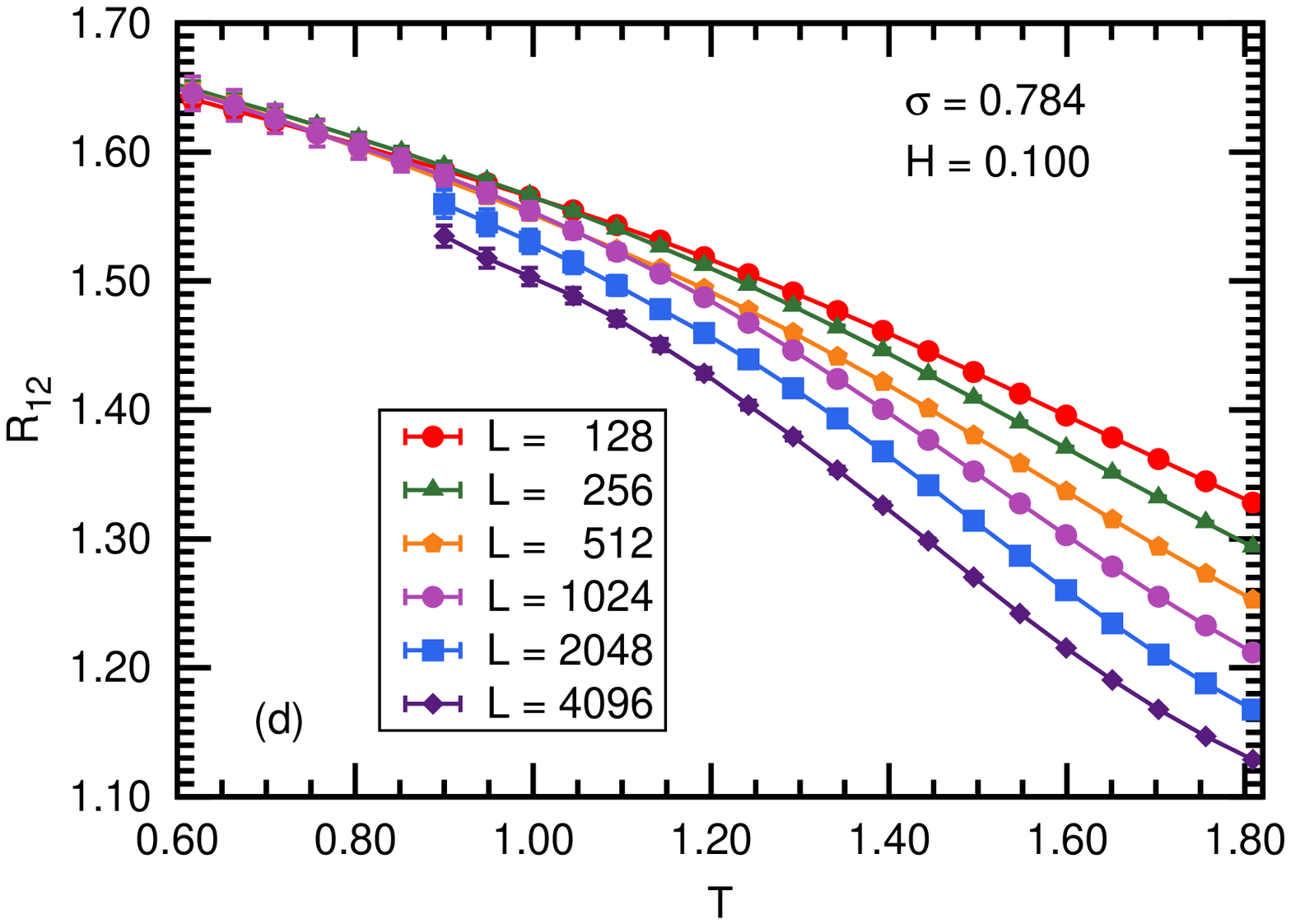}

\vspace*{-0.3cm}
\caption{(Color online)
Data used in the modified FSS analysis for $\sigma = 0.784$. As
in Fig.~\ref{fig:0.784-standard}, the left column is for $H = 0$
and the right column for $H = 0.1$. The top row shows the inverse
spin-glass susceptibility $\chisg(k\to 0)^{-1}$ extrapolated from
results for $k \ne 0$. The bottom row shows data for the dimensionless
ratio $R_{12}$ defined in Eq.~\eqref{eq:R12}. The dashed line for
$\chisg(k\to 0)^{-1}$ [top row, panels (a) and (b)] indicates the
extrapolated transition temperature $T_c$ determined from the fits in
Fig.~\ref{fig:0.784-Tc}. The $R_{12}$ data for $H=0$ [panel (c)] show
intersections but the trend of intersection temperatures is not smooth
for this range of sizes. The dotted vertical line shows the transition
temperature estimated from the intersections of $\chisg/L^{2\sigma-1}$
in Fig.~\ref{fig:0.784-standard} ($T_c=1.36$) and so is just a guide
to the eye. The $R_{12}$ data for $H = 0.1$ [panel (d)] show no sign
of a transition in the range of temperatures studied.
}
\label{fig:0.784-modified}
\end{figure*}

This method requires several nonzero wave vectors and so is particularly
suitable for one-dimensional models as studied here, because one can
simulate very large \textit{linear} sizes for these. For a
\textit{short-range} model in 4D, where the number of wave vectors is
more limited, the authors of Ref.~\onlinecite{banos:12} propose another
quantity. To motivate this quantity, we note as stated above, that
$\xi_L/L$ is particularly useful because it is dimensionless. From
Eq.~\eqref{eq:xiL} we see that the crucial quantity is the ratio
$\chisg(0) / \chisg(k_{1})$, which is also dimensionless. Massaging this
quantity to obtain another dimensionless quantity $\xi_L/L$ according to
Eq.~\eqref{eq:xiL} is actually not essential. Therefore, a related
quantity which does not involve $k=0$ can be defined via
\begin{equation}
R_{12} = {\chisg(k_1) \over \chisg(k_{2})} \, ,
\label{eq:R12}
\end{equation}
where $k_2 = 4 \pi / L$ is the second smallest nonzero wave vector.
Because it is dimensionless, $R_{12}$ has the same FSS form as $\xi_L/L$
shown in Eq.~\eqref{eq:xiscale}. Consequently, curves of $R_{12}$ for
different system sizes $L$ should intersect at $T_c$ if corrections to
FSS are unimportant.

\subsection{Numerical Details}

The simulations are done using the parallel tempering (exchange)
Monte Carlo method.\cite{geyer:91,hukushima:96} Simulation parameters
are listed in Table \ref{tab:simparams}. Equilibration is tested
using the method developed in Ref.~\onlinecite{katzgraber:09b}
[Eq.~(8)]: The energy per spin is computed directly, as well as as
a function of a spin correlator. Both have to agree if a system is
in thermal equilibrium. Starting from a random configuration, the
directly computed energy is typically overestimated, while the energy
computed from the correlator is underestimated. Only when both agree
(on average) is the system in thermal equilibrium. We thus perform a
logarithmic binning of the data and requiring that both the energy per
spin and the energy computer from the correlator agree for at least
the last two logarithmic bins. The reason we wait for two additional
logarithmic bins is because the equality only holds on average. By
being more conservative with the equilibration times we ensure that
the bulk of the samples are in thermal equilibrium.\cite{yucesoy:13}
In addition, we verify that all other observables are independent of
Monte Carlo time for at least these last two bins.

\section{Results}
\label{sec:results}

\subsection{$\boldsymbol{\sigma = 0.784}$ (four space dimensions)}

\begin{figure*}[!tbh]
\center

\includegraphics[width=0.45\textwidth]{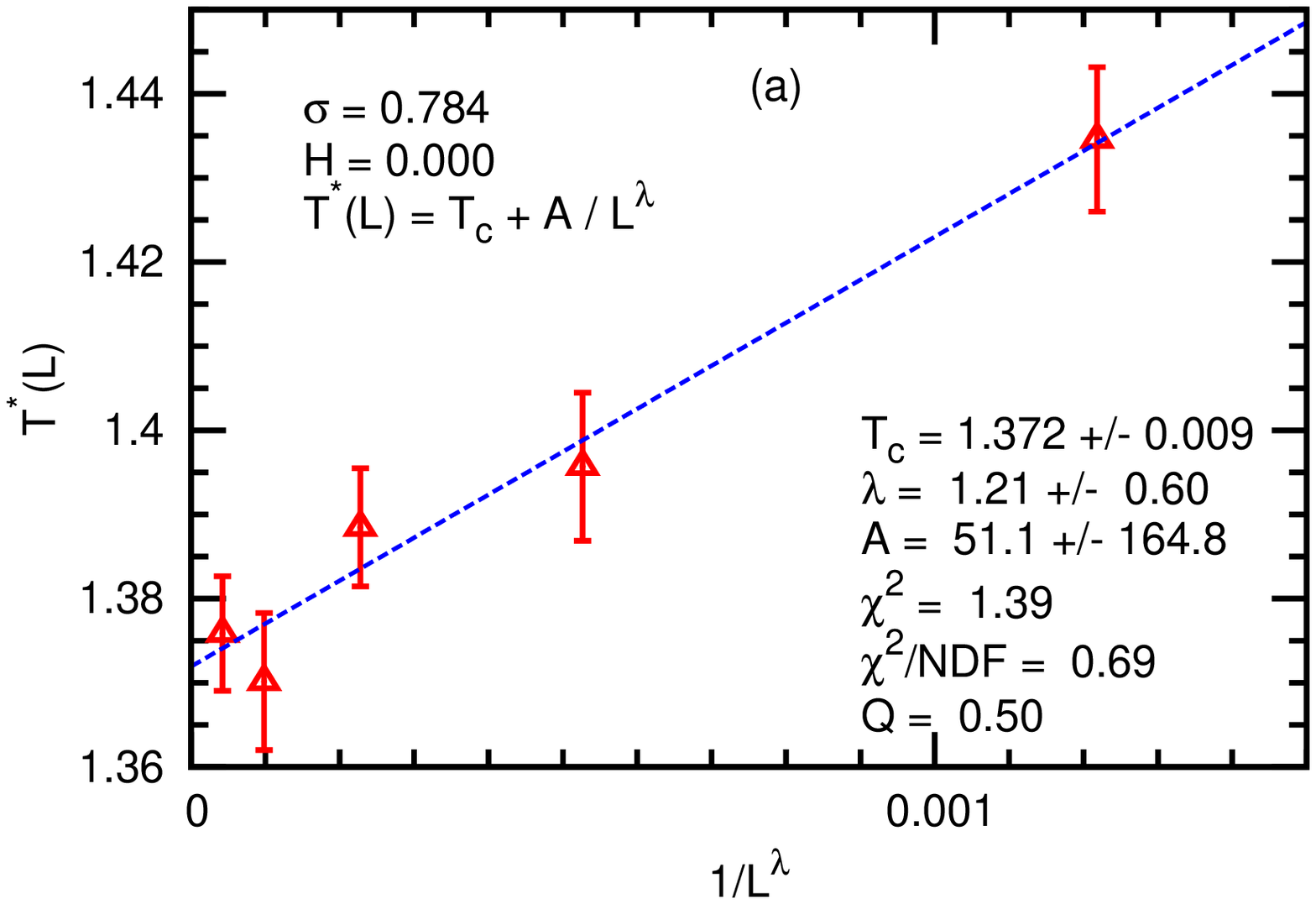}
\hspace*{0.5cm}
\includegraphics[width=0.45\textwidth]{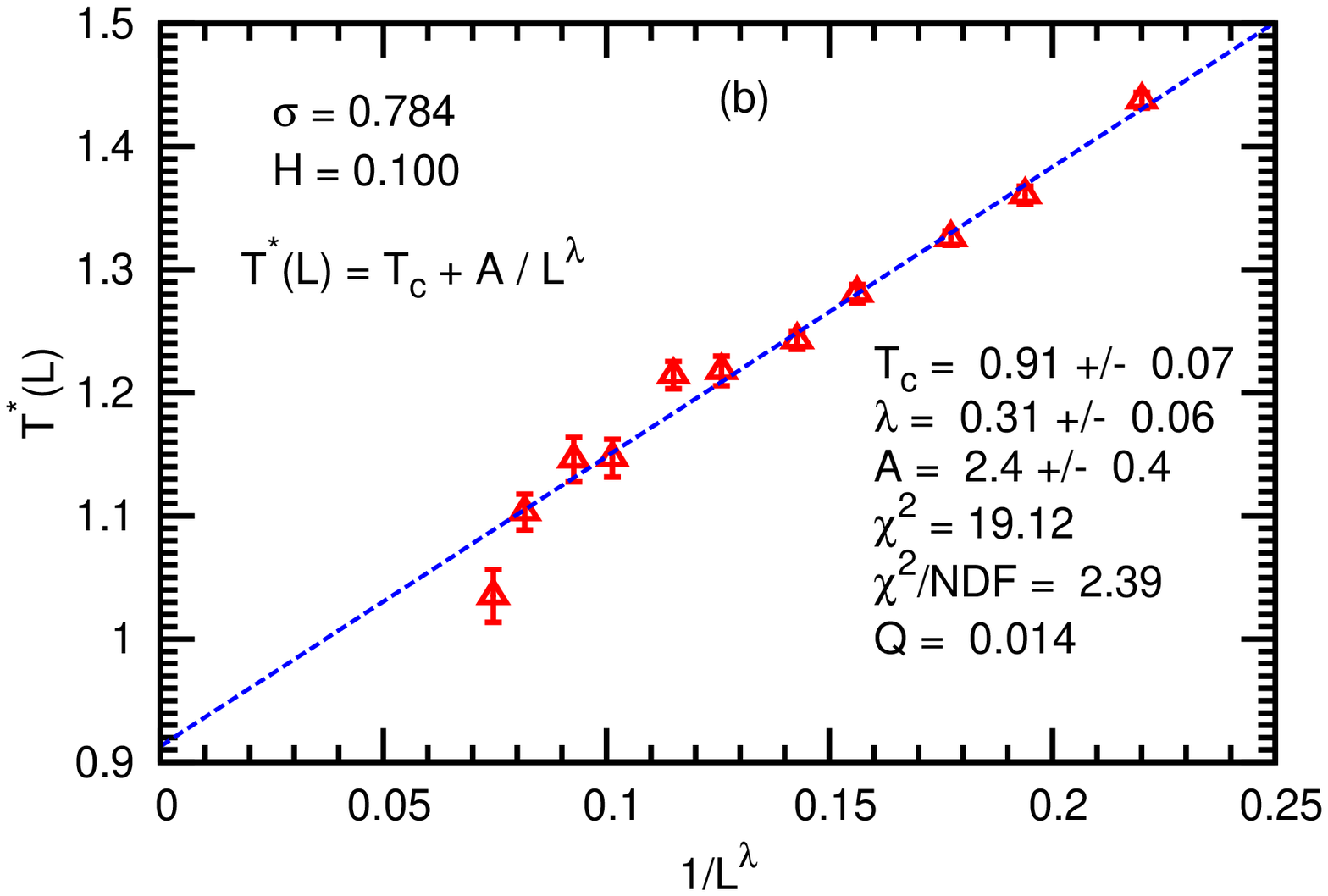}

\vspace*{-0.3cm}
\caption{(Color online)
Values of $T^\star(L)$, the temperature where $\chisg(k\to 0)^{-1}$ goes
through zero, for different system sizes for $H=0$ (left) and $H = 0.1$
(right) for $\sigma = 0.784$. The data are fitted to Eq.~\eqref{eq:Tc}
with $T_c$, $A$, and $\lambda$ as adjustable parameters. In both cases a
finite value of $T_c$ (the intercept) is found. For $H=0$ the 
goodness-of-fit parameter\cite{Q,press:92} $Q$ is not very small, indicating a
satisfactory fit. For $H=0.1$ the value of $Q$ is smaller because the
point for the largest size is well below the fit.
}
\label{fig:0.784-Tc}
\end{figure*}

When $\sigma = 0.784$ the one-dimensional model is a proxy for a
short-range model in four space dimensions. Data for $\xi_L/L$ and
$\chisg/L^{2\sigma-1}$, used within the standard FSS, are shown in
Fig.~\ref{fig:0.784-standard} for $H = 0$ (left column) and $H=0.1$
(right column). The zero-field data for the scaled $\chisg$ in
Fig.~\ref{fig:0.784-standard}(c) show clear intersections indicating a
transition at $T_c \simeq 1.36$, while the data for $\xi_L/L$ in
Fig.~\ref{fig:0.784-standard}(a) show clear intersections at $T_c \simeq
1.34$. This (small) difference is presumably due to corrections to
scaling. Because there is no doubt that there is a zero-field transition
for the models studied in this paper (see e.g.,
Ref.~\onlinecite{banos:12b}) we have not carried out a precise estimate
of the value of $T_c$ in zero field.

In contrast to the zero-field data, the data in a field of $H = 0.1$,
Figs.~\ref{fig:0.784-standard}(b) and \ref{fig:0.784-standard}(d), show
no intersections, indicating the absence of a transition, at least for
this range of temperatures and field.

Data for the extrapolated value of $\chisg(k\to 0)^{-1}$ fitted
according to Eq.~\eqref{eq:parfit} and $R_{12}$ defined via
Eq.~\eqref{eq:R12} are shown in Fig.~\ref{fig:0.784-modified} and used
in the modified FSS. The zero-field data for $R_{12}$ in
Fig.~\ref{fig:0.784-modified}(c) show intersections but the intersection
temperatures do not vary monotonically for this range of sizes. The
dotted vertical line in Fig.~\ref{fig:0.784-modified}(c) corresponds to
$T_c$ from the scaled $\chisg$ data in
Fig.~\eqref{fig:0.784-standard}(c), and so is only a guide to the eye.
The data for $R_{12}$ in a field ($H=0.1$) in
Fig.~\ref{fig:0.784-modified}(d) have no intersections, i.e., no
transition is visible for the range of temperatures studied.

For each system size, the temperature where the data for $\chisg(k\to
0)^{-1}$ shown in Figs.~\ref{fig:0.784-modified}(a) and
~\ref{fig:0.784-modified}(b) goes through zero is referred to as
$T^\star(L)$. This value is obtained by fitting the data to a cubic
polynomial (using the seven data points nearest to zero), and error bars
are obtained from a bootstrap analysis. The results are shown in
Fig.~\ref{fig:0.784-Tc}. The thermodynamic transition temperature $T_c$
is then given by $\lim_{L\to\infty} T^\star(L)$, which we estimate by
fitting to Eq.~\eqref{eq:Tc}. The fits, obtained by adjusting $T_c$, $A$,
and $\lambda$, are shown in the figures (dashed lines). The data
indicate a finite value of $T_c$ both in a field and in zero field. The
central values for $T_c$ are shown by the dashed vertical lines in
Figs.~\ref{fig:0.784-modified}(a) and \ref{fig:0.784-modified}(b). The
value of $\chi^2/N_\text{DOF}$, where $N_\text{DOF}$ is the number of
degrees of freedom and the goodness-of-fit parameter\cite{Q,press:92}
$Q$, indicate a good fit in the case of $H=0$ ($Q = 0.50$). For $H=0.1$,
the value of $Q$ is smaller ($Q = 0.014$) because the point for the
largest size is well below the fit.

Our analysis to ascertain whether there is a finite $T_c$ includes four
sets of data: $\xi_L/L$, $\chisg/L^{2\sigma-1}$, $\chisg(k\to 0)^{-1}$,
and $R_{12}$. In zero field they all clearly show a finite transition
temperature, in agreement with the work of Ba\~nos {\em et
al}.\cite{banos:12b}, who studied almost the same model. However, in a
field of $H=0.1$ there is an inconsistency: Three of the four measures
($\xi_L/L$, $\chisg/L^{2\sigma-1}$, and $R_{12}$) show no sign of a
transition. In contrast, the value of $T_c$ from results for
$\chisg(k\to 0)^{-1}$ does appear to be nonzero ($T_c = 0.91 \pm
0.07$), see Fig.~\ref{fig:0.784-Tc}(b). We note, however, that the error
bar is large and the last data point being below the fit may possibly
indicate a downward trend at larger sizes.

This discrepancy highlights the need to better understand FSS in spin
glasses in a magnetic field. We shall come back to this question in
Sec.~\ref{sec:conclusions}. However, already we note that because the
$\chisg(k\to 0)^{-1}$ data are the \textit{only} indicator for a finite
$T_c$ in a field, we feel we should view results for this quantity with
caution.

\begin{figure*}[!tbh]
\center

\includegraphics[width=0.45\textwidth]{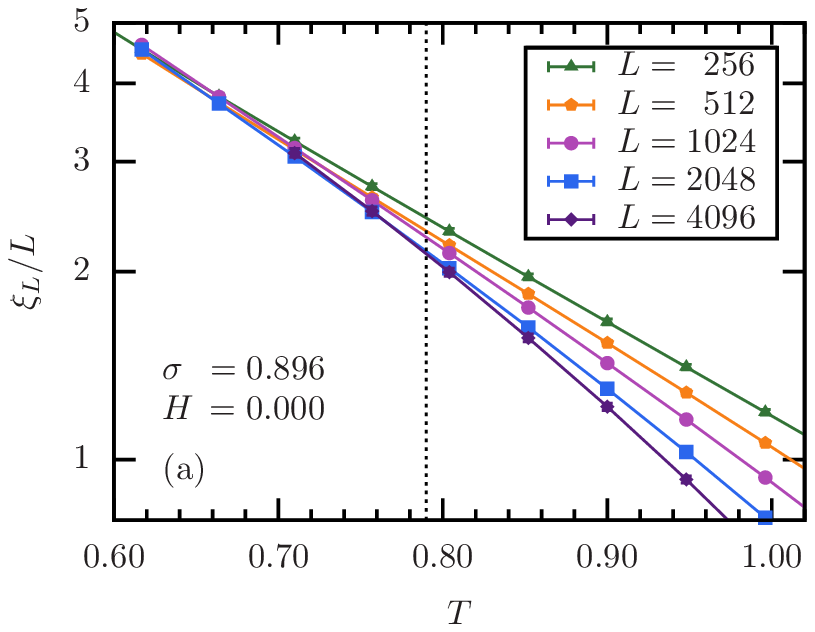}
\hspace*{0.5cm}
\includegraphics[width=0.45\textwidth]{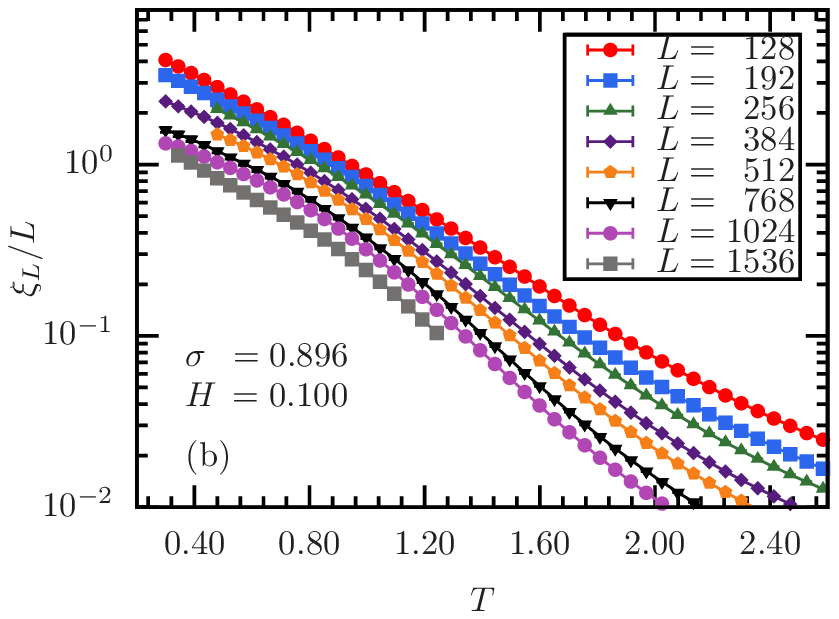}

\vspace*{0.1cm}

\includegraphics[width=0.45\textwidth]{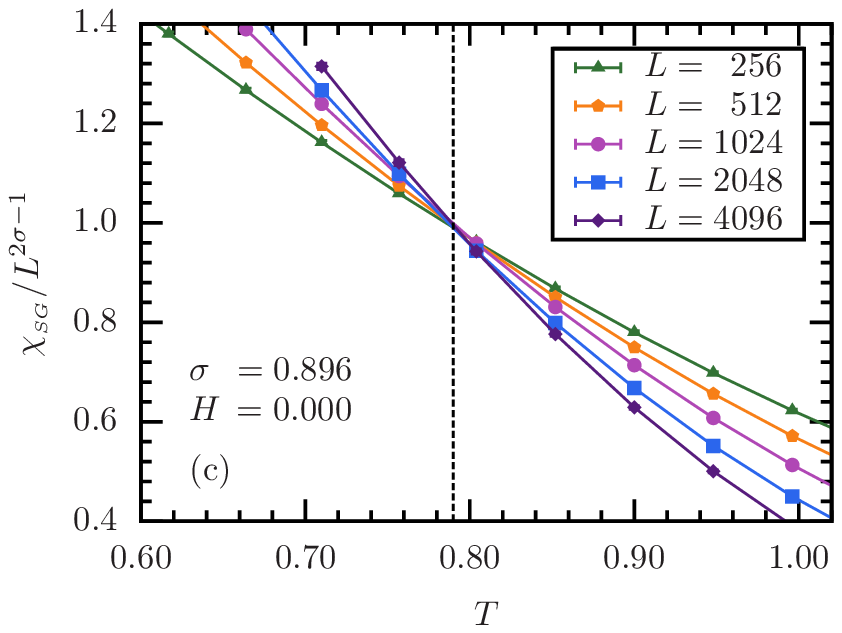}
\hspace*{0.5cm}
\includegraphics[width=0.45\textwidth]{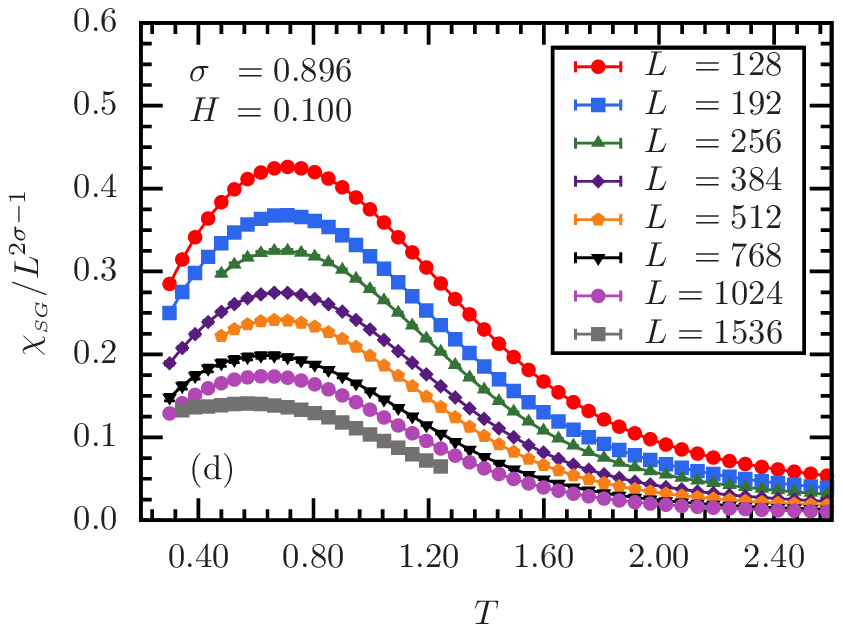}

\vspace*{-0.3cm}

\caption{(Color online)
Data used in the standard FSS for $\sigma = 0.896$, which is a proxy for
three space dimensions. The left column is for $H = 0$ and the right column
for $H = 0.1$. The top row shows the finite-size correlation length
divided by $L$ as a function of $T$ for different system sizes. The
bottom row shows results for the scale-invariant spin-glass
susceptibility, $\chisg/L^{2\sigma-1}$. The intersections in the data
for $\chisg/L^{2\sigma-1}$ for $H=0$ [panel (c)] indicate a transition
at $T_c \simeq 0.795$, which is shown by a dashed vertical line. The
zero-field results for $\xi_L/L$ [panel (a)] show stronger finite-size
effects and the evidence for intersections is not so clear. The same
result was found recently by Ba\~nos {\em et al}.\cite{banos:12b}, who
were able to study somewhat larger sizes for almost the same model.
Their FSS analysis of all their data supports a zero-field transition.
In panel (a) the dotted vertical line indicates the transition
temperature found for the data in panel (c) and so is just a guide to
the eye. The data in panel (a) are consistent with intersections tending
to this value for $L \to \infty$. By contrast, the lack of intersections
in the right-hand column indicates no transition in a magnetic field.
}
\label{fig:0.896-standard}
\end{figure*}

\subsection{$\boldsymbol{\sigma = 0.896}$ (three space dimensions)}

\begin{figure*}[!tbh]
\center

\includegraphics[width=0.45\textwidth]{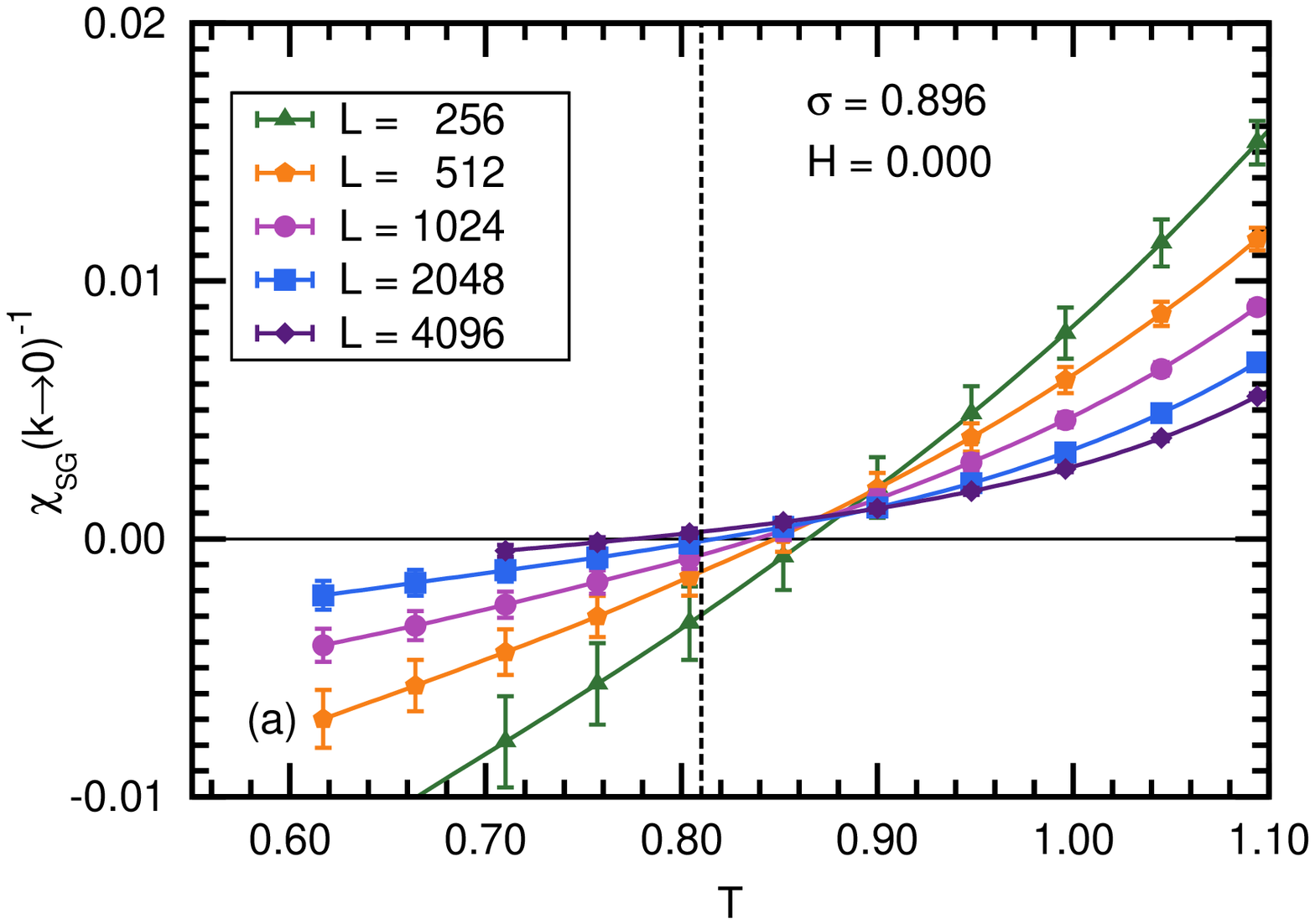}
\hspace*{0.5cm}
\includegraphics[width=0.45\textwidth]{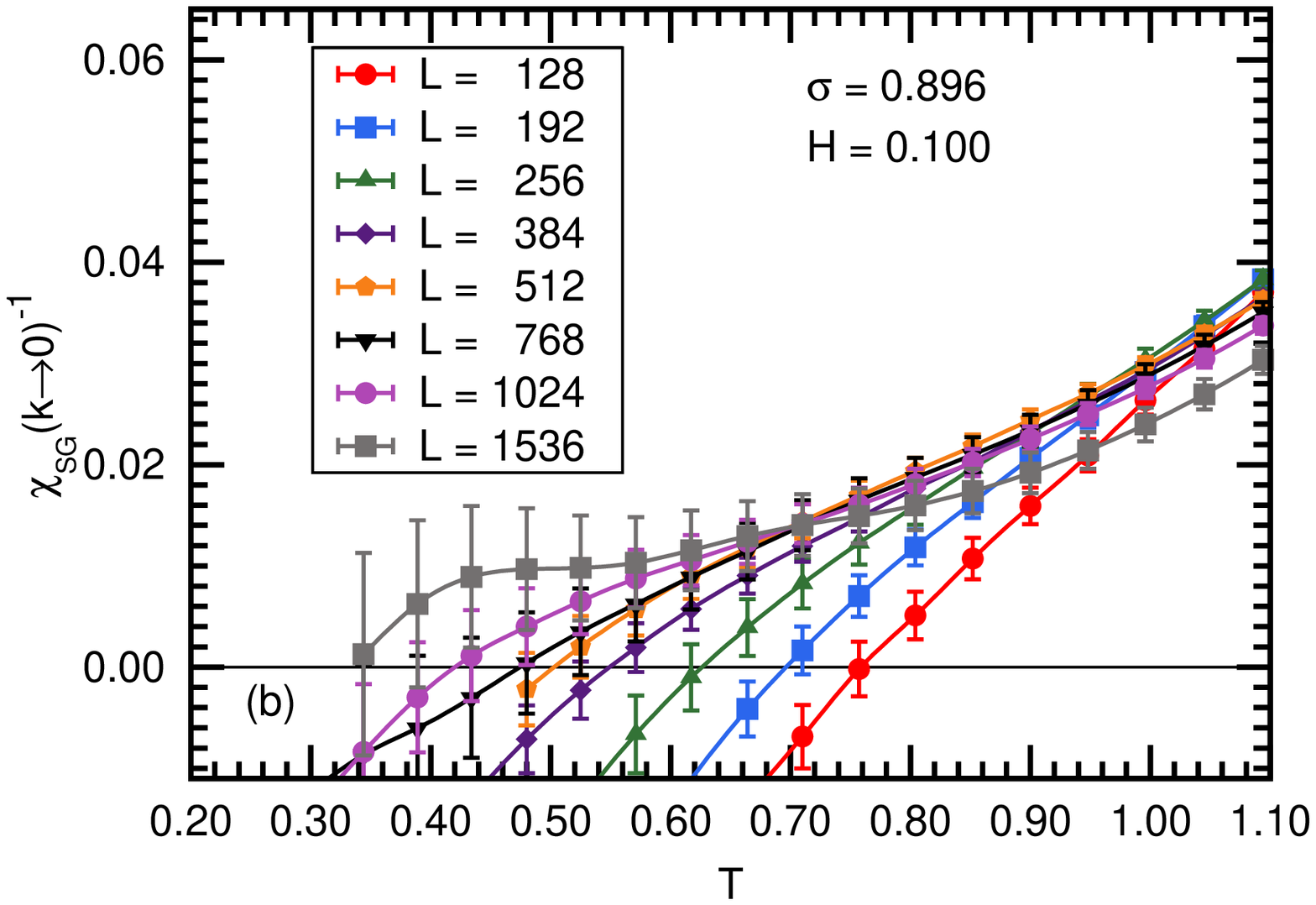}

\vspace*{0.1cm}

\includegraphics[width=0.45\textwidth]{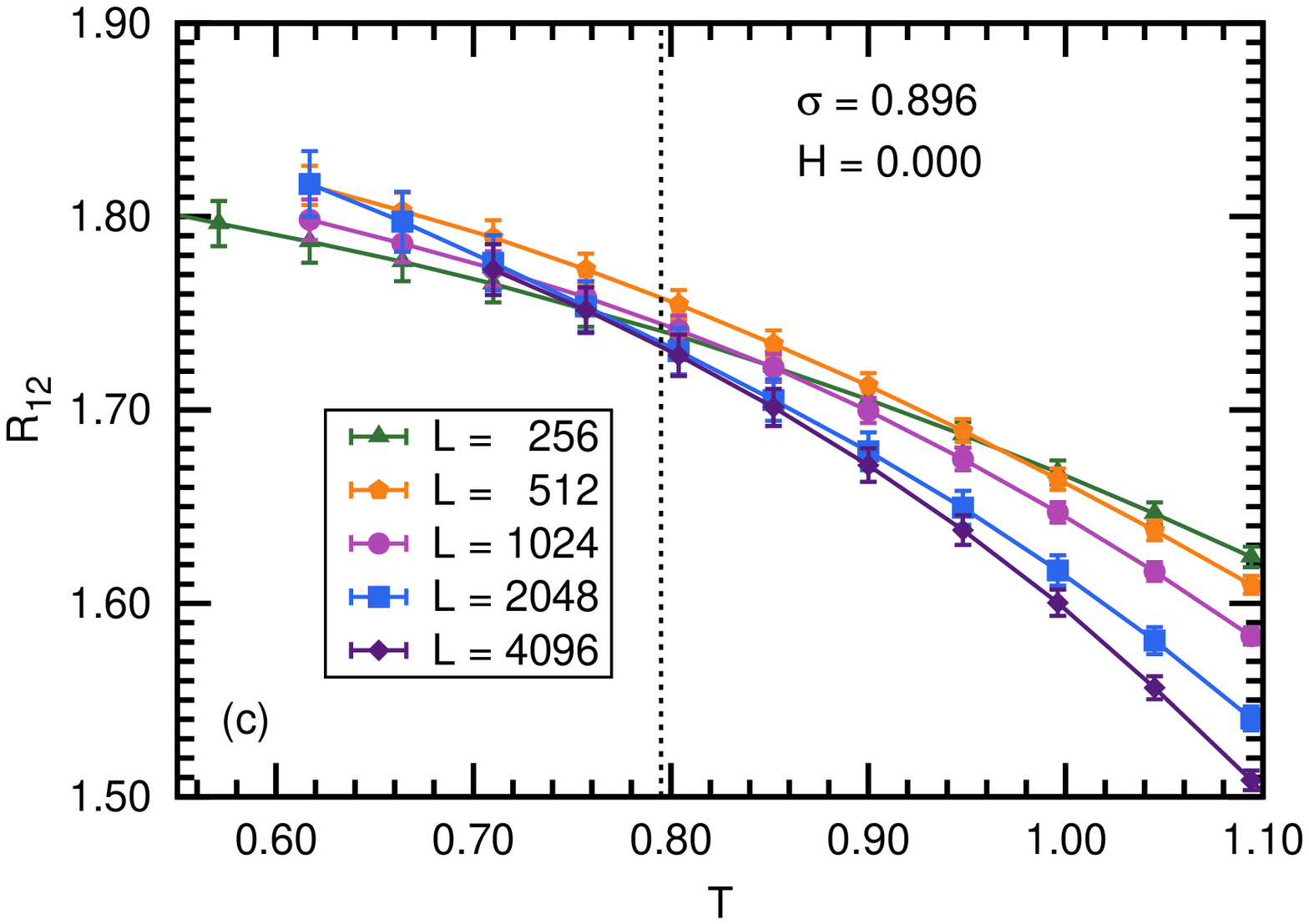}
\hspace*{0.5cm}
\includegraphics[width=0.45\textwidth]{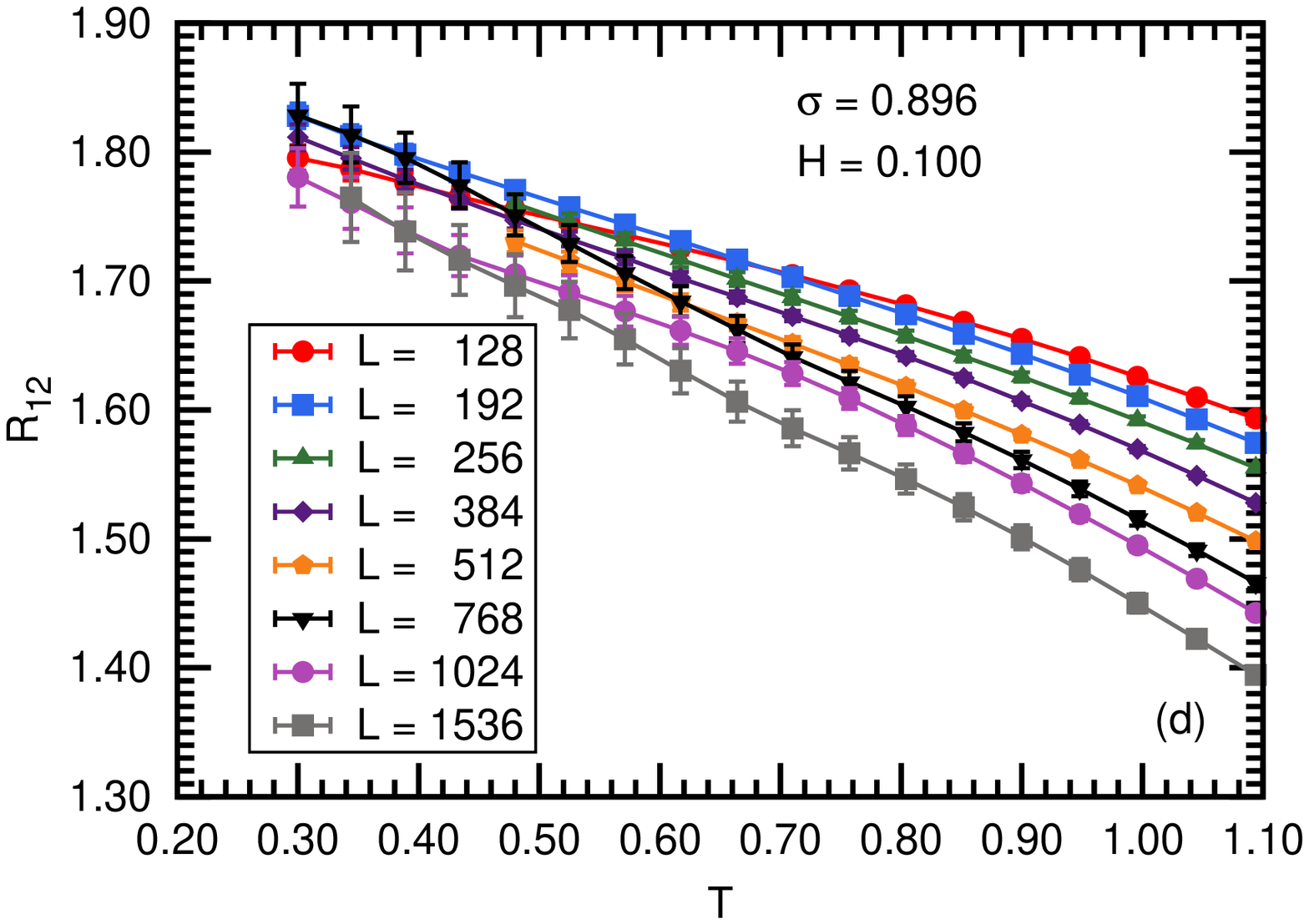}

\vspace*{-0.3cm}

\caption{(Color online)
Data used in the modified FSS analysis for $\sigma = 0.896$. As in
Fig.~\ref{fig:0.896-standard}, the left column is for $H = 0$ and the
right column for $H = 0.1$. The top row shows the inverse spin-glass
susceptibility $\chisg(k \to 0)$ extrapolated from results for $k \ne
0$. The bottom row shows data for the dimensionless ratio $R_{12}$
defined in Eq.~\eqref{eq:R12}. The dashed vertical line in panel (a)
marks the value for $T_c$ obtained by extrapolating the data for
$T^\star(L)$ [where $\chisg(k\to 0)^{-1}$ vanishes] according to
Eq.~\eqref{eq:Tc}; see Fig.~\ref{fig:0.896-Tc}. The result is $T_c =
0.81(3)$, in agreement with the value of $0.795$ from the intersections
of the scaled $\chisg$ in Fig.~\ref{fig:0.896-standard}(c). The data for
$R_{12}$ in zero field in panel (c) show intersections, although the
intersection temperatures do not vary smoothly. The dotted vertical line
shows $T_c$ obtained from the scaled $\chisg$ and is just a guide to the
eye. The results for $R_{12}$ for $H=0.1$ in panel (d) do not provide
clear evidence for a transition.
}
\label{fig:0.896-modified}
\end{figure*}

The one-dimensional long-range model with $\sigma =0.896$ is a proxy for
a short-range spin glass in three space dimensions. The data used in
the standard FSS analysis are shown in Fig.~\ref{fig:0.896-standard}.
Again, the left column is for $H=0$ and the right column is for $H =
0.1$. The zero-field results for the scaled $\chisg$ in
Fig.~\ref{fig:0.896-standard}(c) show a clear transition. The zero-field
data for $\xi_L/L$ in Fig.~\ref{fig:0.896-standard}(a) are less clear
cut because there is little splaying of the data at low
temperatures. However, the temperature where the data merge for two
neighboring sizes \textit{increases} as the system size increases.
Similar results were obtained by Ba\~nos {\em et al}.\cite{banos:12b}
for almost the same model, although they were able to study sizes of up
to $L=8192$ which do show a clear intersection with the data for $L =
4096$ (see Fig.~15 in their paper). Performing a detailed FSS analysis,
Ba\~nos {\em et al}.\cite{banos:12b} showed that all their data
\textit{are} consistent with a finite value of $T_c$. Because our data
for $\xi_L/L$ do not in itself convincingly locate the transition
temperature, the dotted line in Fig.~\ref{fig:0.896-standard}(a) shows
(as a guide to the eye) the location of $T_c$ as determined from the
scaled $\chisg$ data in Fig.~\ref{fig:0.896-standard}(c).

In a field, the data for $\xi_L/L$ in Fig.~\ref{fig:0.896-standard}(b)
and the scaled spin-glass susceptibility in
Fig.~\ref{fig:0.896-standard}(d) show no intersections and, hence,
indicate that there is no transition for this range of temperature and
field.

Data for the modified FSS analysis are shown in
Fig.~\ref{fig:0.896-modified}. The results for $R_{12}$ are shown in the
bottom row. In zero field [see Fig.~\ref{fig:0.896-modified}(c)] there
are clear intersections, although these do not vary monotonically for the
range of sizes studied. However, the data are consistent with the value
$T_c \simeq 0.795$ obtained from the scaled $\chisg$ data in
Fig.~\ref{fig:0.896-standard}(c), and this value is indicated as a guide
to the eye by the dotted vertical line in
Fig.~\ref{fig:0.896-modified}(c). The data for $R_{12}$ in a field
$H=0.1$ [Fig.~\ref{fig:0.896-modified}(d)] do not show clear evidence
for a transition in the range of temperatures studied.

\begin{figure*}[!tbh]
\center

\includegraphics[width=0.45\textwidth]{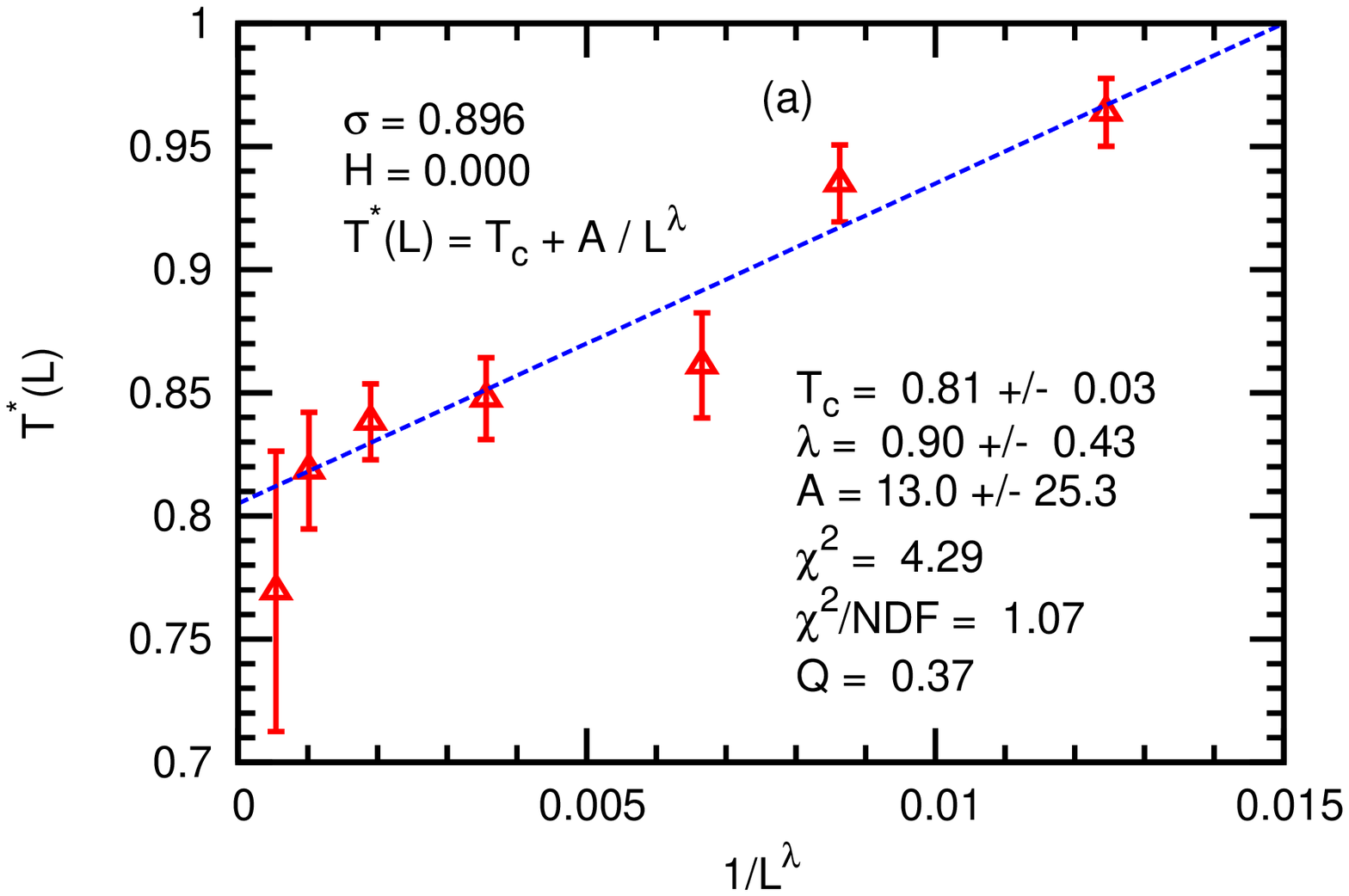}
\hspace*{0.5cm}
\includegraphics[width=0.45\textwidth]{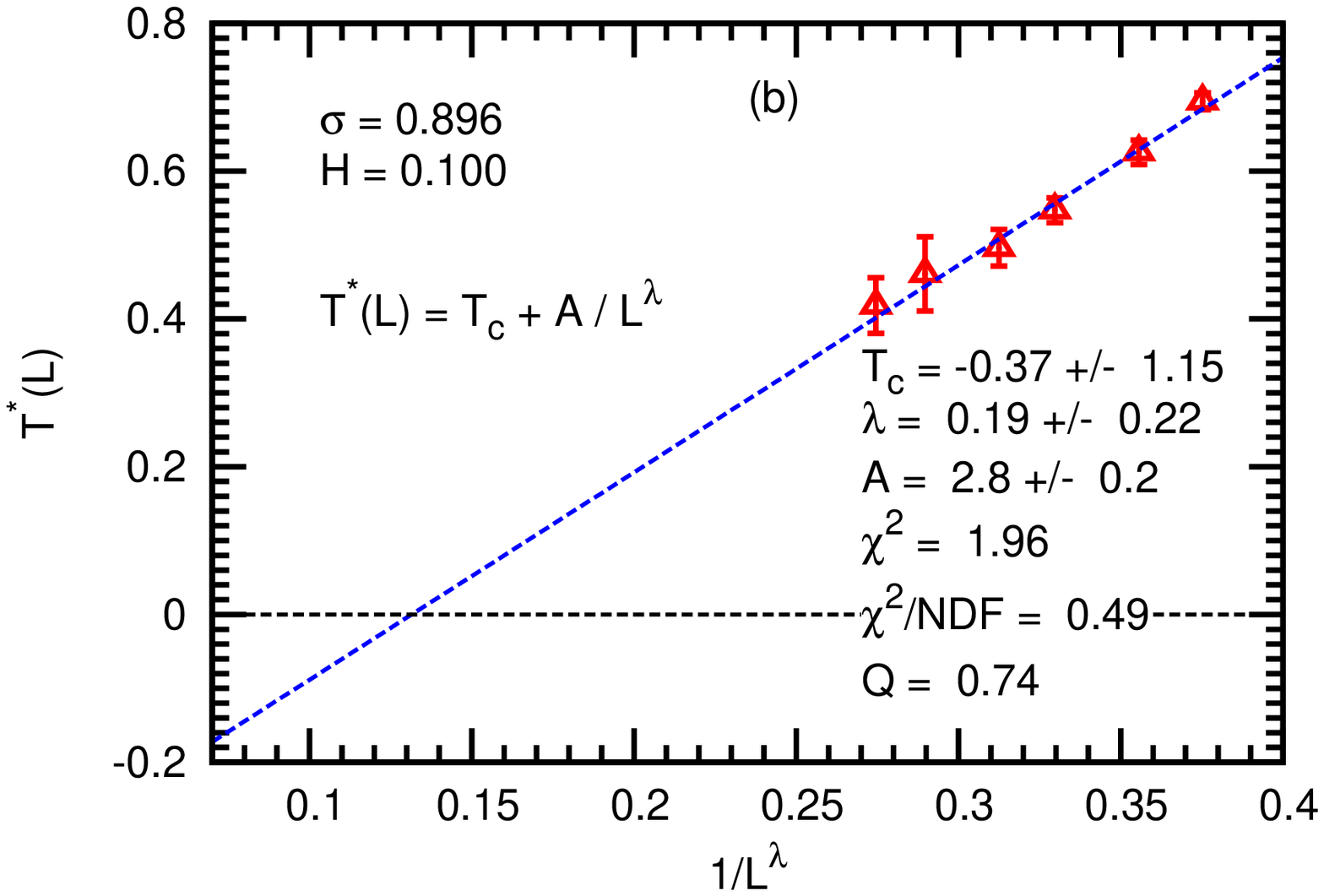}

\vspace*{-0.3cm}
\caption{(Color online)
Values of $T^\star(L)$, the temperature where $\chisg(k\to 0)^{-1}$ goes
through zero, for different system sizes for $H=0$ (left) and $H = 0.1$
(right) for $\sigma = 0.896$. The data are fitted to Eq.~\eqref{eq:Tc}
with $T_c$, $A$, and $\lambda$ as adjustable parameters. The error bars
on the parameters in the fit to the $H=0.1$ data are very large (this is
discussed in the text and in the caption to
Fig.~\ref{fig:cumul_Tc_h0.1_s0.896}). Note that the
$H=0.1$ data are consistent with $T_c = 0$. The goodness-of-fit
parameter\cite{press:92} indicates a satisfactory fit in both cases.
}
\label{fig:0.896-Tc}
\end{figure*}

The top row of Fig.~\ref{fig:0.896-modified} shows results for
$\chisg(k\to 0)^{-1}$. The temperatures where $\chisg(k\to 0)^{-1} = 0$
are plotted and fitted according to Eq.~\eqref{eq:Tc} in
Fig.~\ref{fig:0.896-Tc}. The values of $Q$, $0.37$ for $H=0$ and $0.91$
for $H = 0.1$, indicate a good fit. For $H=0$ the result is $T_c =
0.81\pm 0.03$ and the central value is indicated by the dashed vertical
line in Fig.~\ref{fig:0.896-modified}(a). This value is consistent with
the value $0.795$ from the data for the scaled $\chisg$ in
Fig.~\ref{fig:0.896-standard}(c). For $H = 0.1$, the values of
$T^\star(L)$ shown in Fig.~\ref{fig:0.896-Tc}(b) have a very strong size
dependence. The figure also shows the values of the parameters obtained
by fitting the data to Eq.~\eqref{eq:Tc}. In particular, we find $T_c =
-0.37 \pm 1.15$, indicating that the optimal $T_c$ is negative but the
error bar is very large. This large error bar requires more discussion,
which we now give.

\begin{figure}[!tbh]
\center
\includegraphics[width=0.9\columnwidth]{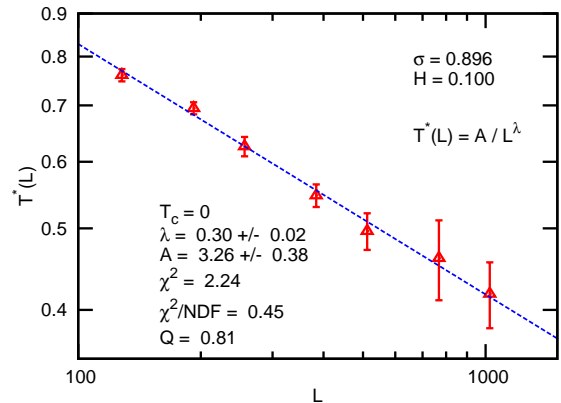}

\vspace*{-0.3cm}
\caption{(Color online)
Log-log plot of the data for $T^\star(L)$ for $\sigma = 0.896$ and $H =
0.1$ assuming $T_c = 0$. The fit works very well according to the
goodness-of-fit parameter $Q$, which is $0.81$.
}
\label{fig:0.896-Tc-0}
\end{figure}

First, we note that a log-log plot of the data in
Fig.~\ref{fig:0.896-Tc-0} indicates that the data are compatible with
$T_c=0$. In addition, we perform the following analysis: For each system
size we construct $200$ bootstrap data sets and estimate $T^\star(L)$
for each of them. There is a huge scatter in the estimates from the
largest size $L=1536$. Thus we ignore this size in the analysis. For
$L=784$, $9$ of the $200$ bootstrap data sets do not yield a temperature
where $\chi_{\rm SG}(k\to 0)^{-1}$ vanishes. Hence, we consider $191$
bootstrap data sets and fit each of them according to Eq.~\eqref{eq:Tc}.

\begin{figure}[!tbh]
\center
\includegraphics[width=0.9\columnwidth]{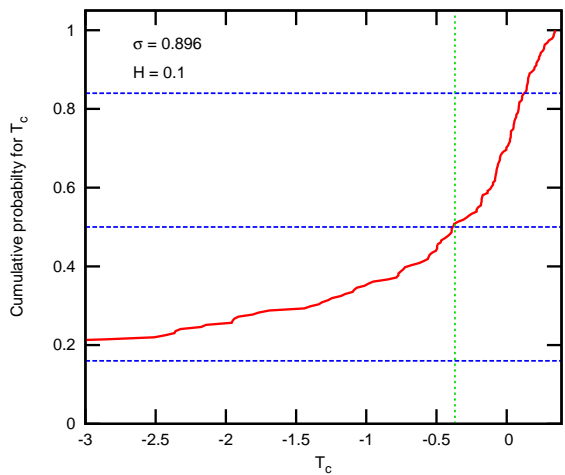}

\vspace*{-0.3cm}
\caption{(Color online)
Cumulative distribution of the transition temperature $T_c$ obtained
from a bootstrap analysis of the data for $\sigma = 0.896$ and $H=0.1$.
The estimate of $T_c$ from the global fit ($T_c = -0.37$, indicated by
the dashed vertical line) agrees well with the median (50th percentile) of
the bootstrap estimates, which is indicated by a horizontal dashed line.
Also indicated by horizontal dashed lines are the 16th and 84th percentiles
which would correspond to one standard deviation if the distribution of
$T_c$'s were Gaussian, which is clearly not the case. Only 30\% of the
bootstrap fits have a positive $T_c$.
}
\label{fig:cumul_Tc_h0.1_s0.896}
\end{figure}

Figure \ref{fig:cumul_Tc_h0.1_s0.896} shows the resulting cumulative
distribution of transition temperatures, i.e., the probability that the
transition temperature is less than the stated value. We find that $34$
of the $191$ data sets do not have a minimum in $\chi^2$; rather,
considering $\chi^2$ as a function of $T_c$ while optimizing with
respect to the other fit parameters, $\chi^2$ decreases monotonically as
$T_c \to -\infty$ while $\lambda \to 0$ in this limit. The estimate of
$T_c$ from the global fit ($T_c = -0.37$, indicated by the dashed vertical
line) agrees well with the median (50th percentile) of the bootstrap
estimates. The median is indicated by a horizontal dashed line. Also
indicated by horizontal dashed lines are the 16th and 84th percentiles, which
would correspond to one standard deviation if the distribution of
$T_c$'s were Gaussian, which is clearly not the case here. Only 30\% of
the bootstrap fits have a positive $T_c$. We therefore conclude that a
positive $T_c$ is somewhat unlikely but cannot be completely excluded
by the data for $T^\star(L)$.

To conclude this subsection, \textit{all} data are consistent with there
being no AT line at $H = 0.1$ for $\sigma = 0.896$. The results for
$T^\star(L)$ obtained from the vanishing of $\chisg(k \to 0)^{-1}$ do
not exclude a finite $T_c$, but this possibility does not seem to be
supported by the rest of the data.

\section{Conclusions} 

\label{sec:conclusions}

We have presented results of simulations of one-dimensional spin-glass
models in a magnetic field which are proxies for short-range models in
three and four space dimensions. We have analyzed the results using both
a traditional FSS approach (which uses $k=0$ data) and a
recently proposed modified FSS approach which uses only $k > 0$ data.

For the model which is a proxy for a 3D system, all our results are
consistent with there being no transition in a magnetic field, at least
for the range of fields and temperatures that we can study. The results
for $T^\star(L)$, obtained from the modified FSS analysis, are just
compatible with a finite $T_c$ but the other data are only compatible
with the absence of a transition, at least assuming that the data are in
the asymptotic scaling region.

For the model which is a proxy for a 4D system, three of the
four sets of data indicate the absence of a transition in a field
($\xi_L/L$, $\chisg/L^{2\sigma -1}$, and $R_{12}$), while that for
$T^\star(L)$ gives a satisfactory fit indicating $T_c > 0$. This
contradiction indicates that at least some of the data cannot be in
the asymptotic scaling region.

It is, therefore, crucial to understand whether it is better to use $k=0$
data in the analysis as in the standard approach or to exclude that
data as in the modified approach.\cite{leuzzi:09,leuzzi:11,banos:12}
References \onlinecite{leuzzi:09}, \onlinecite{leuzzi:11}, and
\onlinecite{banos:12} argue that the $k=0$ data have strong corrections
to FSS. On the other hand, the divergence occurs at $k=0$ and normally
one uses divergent quantities in FSS because these should show the
asymptotic FSS behavior for the smaller system sizes. It should also be
noted that Ref.~\onlinecite{joerg:08a} found good agreement for the
location of the AT line for the spin glass on a random graph, i.e., the
Viana-Bray model (which corresponds to the $\sigma = 0$ limit of the
present model), using the scaled $\chisg$, i.e., the $k=0$ fluctuations.

Finally, an alternate interpretation of the results can be done using
droplet scaling
arguments.\cite{fisher:87,fisher:88,bray:86,mcmillan:84a} The size
$\xi_{\rm D}$ of the droplets within this picture can be estimated by
equating the domain-wall energy required to create them, $\sim J
\xi_{\rm D}^{\theta}$, to the energy which can be gained from flipping a
droplet of size $\xi_{\rm D}$ in the field, $\sim H \xi_{\rm D}^{d/2}$,
where $d$ is the space dimension.  For the long-range models studied
here, $\theta =1-\sigma$.\cite{katzgraber:03} Thus, the droplet size is
of order $3320$ for $\sigma = 0.784$ and is of order $335$ at $\sigma
=0.896$, when $H/J=0.1$, the ratio used in this study. We have only one
data point, that for $L= 4096$ in Fig.~\ref{fig:0.896-Tc}(b), greater
than $3320$. Interestingly, it is the data points at the smaller values
of $L$ which point to a finite value of $T_c$. The last point at system
size $4096$ lies well below the fitted curve. This suggests that had we
been able to obtain data for a range of system sizes significantly
greater than $4096$, it might have been possible to obtain results in
the analog of four dimensions like that displayed in
Fig.~\ref{fig:0.896-Tc-0} for the analog of three dimensions, where
the last five data points are all greater than the estimated droplet
size there of $335$ and the extrapolated value of $T_c$ is zero.

Ideally, one would determine which set of data is in the asymptotic
scaling regime by simulating larger system sizes. However, because the
present study involved a rather substantial amount of CPU time, this is
not feasible for us at present. Based on the data shown, the balance of
the evidence is that there is no AT line in the one-dimensional models
which are proxies for three and four dimensional short-range spin
glasses. However, a deeper insight into corrections to FSS in spin
glasses is needed to confirm this conclusion.

\begin{acknowledgments} 

H.G.K.~acknowledges support from the Swiss National Science Foundation
(Grant No.~PP002-114713) and the National Science Foundation (Grant
No.~DMR-1151387). We would like to thank the Texas Advanced Computing
Center (TACC) at The University of Texas at Austin for providing HPC
resources (ranger and lonestar clusters), ETH Zurich for CPU time on the
brutus cluster, and Texas A\&M University for access to their eos and
lonestar clusters. A.P.Y acknowledges support from the NSF under Grants
No.~DMR-0906366 and No.~DMR-1207036.

\end{acknowledgments}

\bibliography{refs,notes}

\begin{thebibliography}{40}
\bibitem[{Q()}]{Q}
\bibinfo{note}{The goodness-of-fit parameter $Q$ is the probability that, given
  the fit, the data could have the specified value for $\chi^2$ or greater,
  assuming Gaussian noise; see, for example, Ref.~\onlinecite{press:92}.}
\end{thebibliography}
 
\end{document}